# Advancing Atom Probe Tomography of SrTiO$_3$: Measurement Methodology and Impurity Detection Limits

J.E. Rybak[1], J. Arlt[1], B. Gault[2,3], C.A. Volkert[1]

*1 Institute of Materials Physics, University of Göttingen, Friedrich-Hund-Platz 1, 37077 Göttingen, Germany*
*2 Max Planck Institute for Sustainable Materials, Max-Planck-Straße 1, 40237 Düsseldorf, Germany*
*3 Department of Materials, Imperial College London, Kensington, London, SW7 2AZ, United Kingdom*

**J. E. Rybak**

University of Göttingen, Institute of Materials Physics

Friedrich-Hund-Platz 1, D-37077 Göttingen, Germany

E-Mail: janerik.rybak@uni-goettingen.de

Phone: +49 551 39 26972

**J. Arlt**

University of Göttingen, Institute of Materials Physics

Friedrich-Hund-Platz 1, D-37077 Göttingen, Germany

E-Mail: jarlt@gwdg.de

**Dr. B. Gault**

Max-Planck-Institute for Sustainable Materials, Department of Microstructure Physics and Alloy Design

Max-Planck-Straße 1, D-40237 Düsseldorf, Germany

E-Mail: b.gault@mpie.de

**Prof. C. A. Volkert (corresponding author)**

University of Göttingen, Institute of Materials Physics

Friedrich-Hund-Platz 1, D-37077 Göttingen, Germany

E-Mail: cvolker@gwdg.de

Phone: +49 551 39 25011

**Conflict of interest: The authors declare none**.

Advancing Atom Probe Tomography of SrTiO$_3$

**Abstract:** Strontium titanate (STO) possesses promising properties for applications in thermoelectricity, catalysis, fuel cells, and more, but its performance is highly dependent on stoichiometry and impurity levels. While atom probe tomography (APT) can provide detailed three-dimensional atomic-scale chemical information, STO specimens have been challenging to analyze due to premature specimen fracture. In this study, we show that by applying a thin metal coating to atom probe tips, STO specimens can be analyzed with nearly 100% success. Using this approach, we investigate both undoped STO and 1 at% Nb-doped STO, achieving sufficient sensitivity to detect Nb concentrations as low as 0.7 at%. This work establishes a reliable APT method for high-resolution chemical analysis of STO at the nanoscale.

**Keywords:** Strontium titanate, Atom Probe Tomography, Detection Limit, Tip Fracture

**Graphical Abstract:**

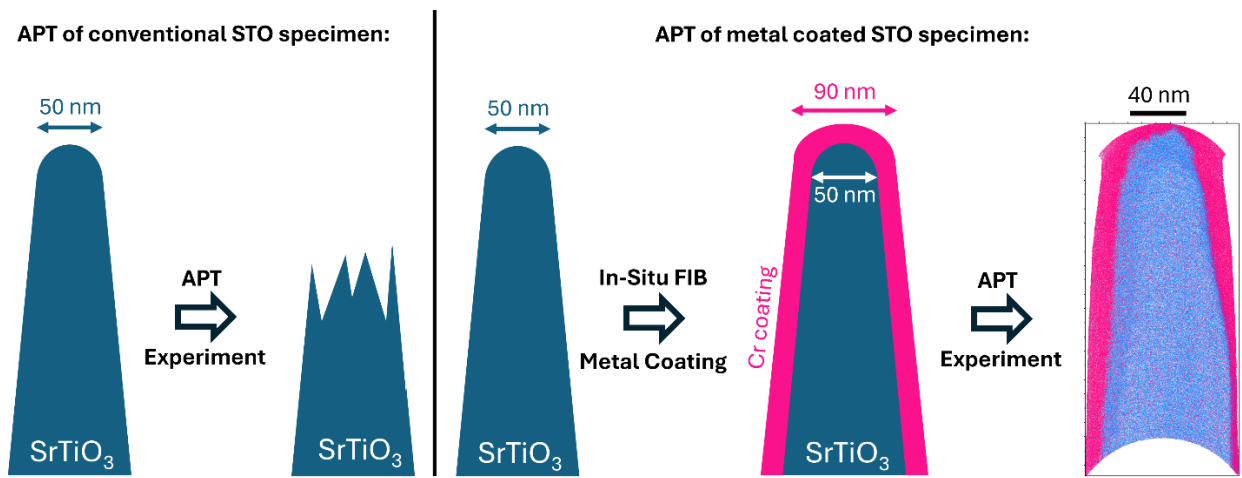

## Introduction:

Oxide perovskites have been extensively studied due to their potential use in a wide range of applications including thermoelectricity (Somaily et al., 2017), catalysis (Zhu et al., 2015), fuel cells (Zhou et al., 2016), and solar cells (Yin et al., 2019). However, their properties are highly sensitive to defect structure, stoichiometry and impurities (Schlom et al., 2008; Mannhart & Schlom, 2010; Zubko et al., 2011), making property control difficult but also providing opportunities for optimization. Examples for SrTiO$_3$ include the role of stoichiometric point defects in controlling transport and optical properties (Chan et al., 1981; Lee et al., 2018), the impact of grain boundaries on electronic and ionic transport (Waser, 1995), the role of surface states on photocatalytic activity (Pattanayak et al., 2022), and possible atomic scale origins for the colossal conductivity observed at some hetero-interfaces (Ohtomo & Hwang, 2004). Thus, understanding the distributions of atoms and impurities of oxide perovskites is crucial for optimizing them for specific applications.

Atom probe tomography (APT) offers a combination of sub-nm 3D spatial resolution and chemical sensitivity down to the ppm-level with a similar sensitivity across the entire periodic table. This makes APT a unique technique for the chemical analysis of features such as grain boundaries and the spatial distribution of trace and light elements and offers an ideal complement to electron microscopy methods (Vurpillot et al., 2016; Herbig, 2018; Gault et al., 2021). However, conventional APT of perovskites like SrTiO$_3$ (STO) and BaTiO$_3$ (BTO) is difficult. Measurements of highly Nb doped bulk STO appear to be successful (Rodenbücher et al., 2013, 2016) and show a homogeneous Nb distribution and FIB-induced Sr enrichment at the specimen apex. However, these results could not be replicated, and a lack of detail makes assessment of the measurement method difficult. Subsequent measurements of STO thin films (Poplawsky et al., 2023; Morris et al., 2024) as well as BTO bulk specimen (Jang et al., 2024) and nanoparticles (Kim et al., 2023) using conventionally prepared specimens have reported success rates between 10% and 20%. These specimens tended to fracture before significant amounts of data could be collected. Presumably, the brittle STO and BTO specimens could not withstand the mechanical stresses induced by the strong electrostatic field necessary to field evaporate the atoms during atom probe analysis.

Several methods have been proposed to improve the yield of APT measurements of brittle and insulating materials such as the oxide perovskites. For instance, it has been shown that a shorter laser wavelength can be beneficial to the APT analysis of STO (Poplawsky et al., 2023). Measurements of a multilayer system containing STO-layers were performed with a 266 nm wavelength laser on a Cameca LEAP 6000XR without the specimens breaking and a larger dataset was obtained as a result. Furthermore, artificial intermixing at the interfaces between layers, which was observed in datasets collected with a 355 nm laser, was suppressed with the 266 nm laser. However, atom probes equipped with such lasers are not yet widely available. An alternative involves coating specimen with a 10 - 20 nm thick conformal metal capping layer (Schwarz et al., 2024). This has been shown to improve the success rates for APT measurements of insulating materials in general and for BTO specifically (Jang et al., 2024). Similarly, embedding BTO nanoparticles in a Ni matrix has a positive effect, provided that enough of the final APT specimen is made up of the Ni matrix (Kim et al., 2023). The metal layer is thought to strengthen the specimen mechanically and shield the core of the specimen from the electric field, thus reducing the stresses induced by the field (Kim et al., 2022; Schwarz et al., 2024).

Here, we have used the approach of applying a metal coating to APT specimens of STO and show that it significantly increases the measurement yield. Undoped and Nb-doped STO crystals have been analyzed and the composition and Nb distribution are found to be uniform and similar across all specimens. From the Nb doped specimen we derive the detection limit for Nb as a measure of the spatial and concentration resolution for impurity element distributions and discuss the implications for detecting local concentration variations at defects.

## **Materials and Methods:**

APT was performed on specimens extracted from two commercially sourced (100) out-of-plane STO wafers: one undoped and one doped with 1at% Nb. To prevent charging effects during FIB preparation, the undoped wafers were deposited with a 100 nm thick layer of amorphous, Nb doped STO by means of ion beam sputtering at room temperature with a partial oxygen pressure of $1.8 \cdot 10^{-4}$ mbar using a commercial STO target.

APT specimens were machined from the near surface regions of the wafers using a Thermo Fisher Scientific Helios G4 dual beam SEM/FIB instrument via a conventional lift-out procedure. A micrometer size block of STO material was cut from the surface and attached to a W support stub using FIB-induced Pt deposition. Afterwards, the STO block was sharpened into a cone shaped tip with an apex radius between 15 and 30 nm using lateral FIB cuts from the side (Miller et al., 2007). The shank angles were held constant between 3° to 4° for all APT specimens. After cleanup with a 5kV, 15pA ion beam from above, the sharpened specimens were coated with a thin (10 to 30 nm) layer of either Cr, Co, Ni or W using redeposition from milling a target of the desired coating metal inside the FIB (Schwarz et al., 2024).

The APT measurements were performed in a custom-built laser assisted wide angle atom probe with a 133 mm straight flight path and a conventional large diameter electrode, for details see (Maier et al., 2016). The system was operated at a wavelength of $\lambda$ = 355 nm with a pulse length of 15 ps, a focus spot size of approximately $\omega_{1/2}$ = 150 μm and a repetition rate of 100 kHz. The base temperature was varied between 70 and 120 K and the laser pulse energy (LPE) was varied between 0.06 and 0.145 μJ. The detection rate window was set to 0.15 - 0.3 ions per 100 pulses. The data reconstruction and analysis were performed with the Scito 2.3.6 software from Inspico (Stender & Balla, n.d.) Further analysis was performed with the open-source software 3depict 0.0.23 (Haley & Ceguerra, 2020) and custom python scripts. The reconstruction was performed with the standard point-projection protocol of Geiser et al. (Geiser et al., 2009) with values for the shank angle and initial radius obtained from transmission electron microscopy (TEM) and, where possible, refined by matching the dimensions of the STO core in the reconstructed APT-dataset to TEM images. Standard deviations for the composition analysis have been calculated according to Danoix et al. (Danoix et al., 2007a, 2007b)

TEM images were taken on a FEI Tecnai G2 Spirit and FEI Titan 80-300 E-TEM.

## Results:

APT measurements of metal coated and uncoated STO were performed for different base temperatures, laser pulse energies, and various coating metals. The experimental conditions and total number of registered counts before end-of-measurement are listed in Table S1. The influence of the metal coatings and measurement parameters on the success rate are described here, and a dataset obtained under optimized measurement conditions is analyzed.

### Experimental Conditions for Optimizing Specimen Yield

The most decisive parameter affecting measurement success is the presence of a suitable metal coating on the STO tip. Without coating, all but one of the STO specimens failed by fracture after fewer than $10^5$ counts, independent of the STO doping, laser parameters and base temperature. This yield is similar to what is reported for uncoated STO in the literature (Poplawsky et al., 2023; Morris et al., 2024). However, after applying a 10-30 nm thick metal coating to the APT specimens, longer runs became possible. Different metallic coatings of Cr, Co, Ni and W were tested to see their influence on the APT measurements. All coatings could be deposited as thin, continuous polycrystalline films as exemplified in the TEM images of a Cr coated specimen in Figure 1. The Cr, Co and Ni coatings prevented early specimen fracture and no significant difference in success rate between the three coatings was observed. In contrast, specimens with W coatings suffered mechanical failure before a significant amount of Sr, Ti and O was collected.

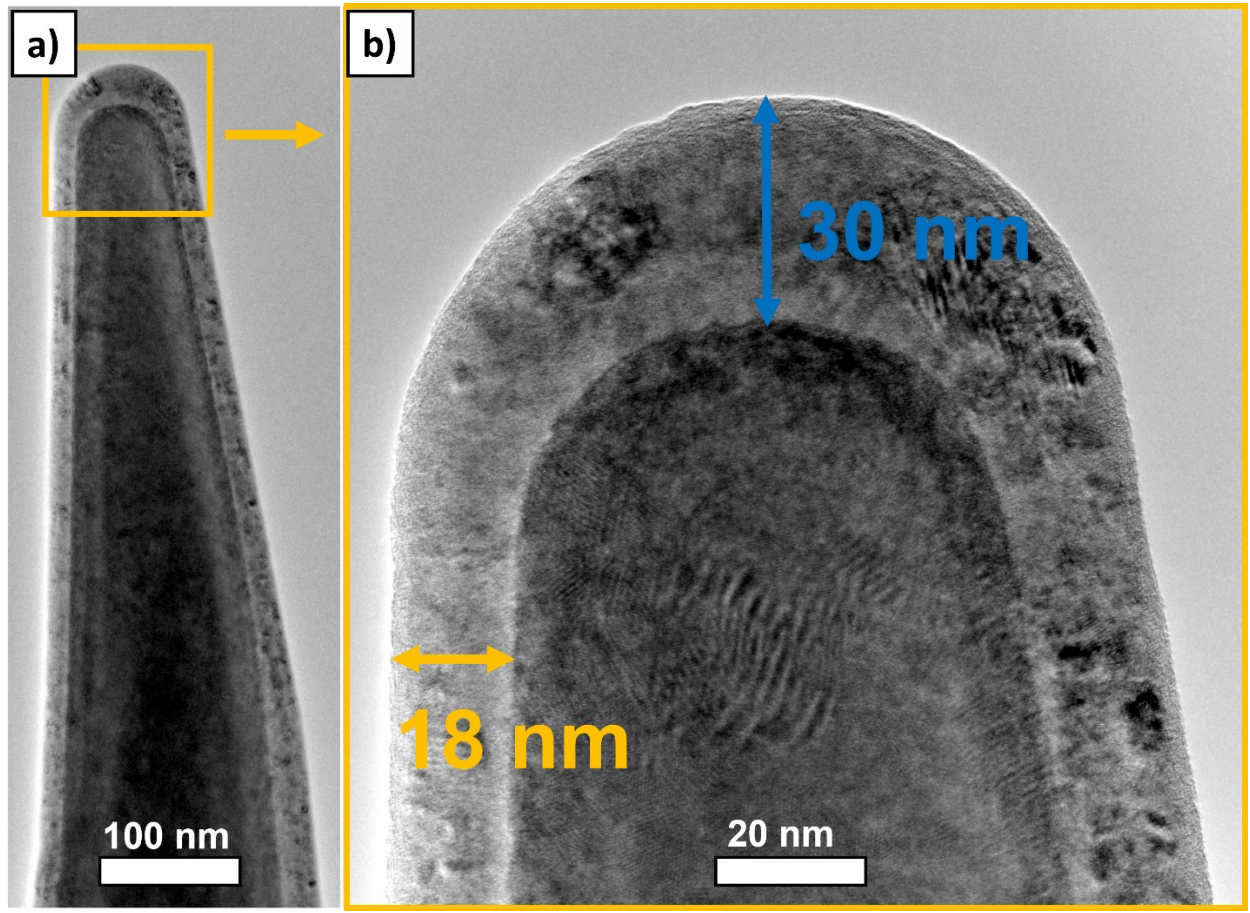

**Fig. 1.** *TEM images of a Cr coated STO specimen a) Overview image of the specimen. The coating has a smooth surface and extends far down the specimen. b) Close-up of the specimen apex.*

The base temperature, target detection rate and laser pulse energy were varied during the measurement runs, but only the base temperature and detection rate were found to have a significant effect on the success rate of coated STO specimens (Table S1). Measurements of Cr coated specimens at 60 and 70 K base temperatures fractured after around $4 \cdot 10^5$ counts of Sr, Ti and O had been collected from the core region of the specimen. Increasing the base temperature to above 115 K allowed for the collection of much larger datasets, typically above $2 \cdot 10^6$ atoms of Sr, Ti and O. The increase in base temperature by ~50 K has no discernible impact on the mass spectrum and no increase in thermal tailing or multi hits beyond the run-to-run variations. Similarly, the increase in base temperature had no impact on the uniformity of the field evaporation as seen in the detector hit map and the homogeneity of the distribution of Sr, Ti, or O. All successful measurements were performed with a low target detection rate of 0.15 - 0.3 ions per 100 pulses; tests with higher target detection rates were unsuccessful with early termination from specimen fracture. The Nb doping and by extension the intrinsic conductivity of the specimen had no observable effect on the measurement success. In general, the STO background count level was an order of magnitude higher than the background for measurements of Al, Fe and W, which we attribute to the low detection rates and relatively high temperatures needed for STO measurements.

Using optimized measurement parameters, the coated STO specimens ran successfully in the atom probe until either the experiment was ended, or the maximum voltage was reached. However, although they did not fail catastrophically, many of them exhibited one or more non-catastrophic fracture events during APT measurement, usually referred to as micro-fractures and common for brittle materials (Gault et al., 2012). These events manifested as spikes in the detection rate, causing the control software to abruptly lower the base voltage (Figure 2a). Afterwards, the software slowly raised the voltage, until the detection rate returned to the target window, usually at a slightly higher voltage than prior to the micro-fracture event (see the voltage curve in Figure 2a). During the recovery period after the micro-fracture, more ions from the metal coating and fewer from the STO core are detected (see the curves for Cr, Sr, and Ti in Figure 2a), before slowly returning to steady state values. Figure 2b shows a TEM image of a specimen, for which the measurement was interrupted during this steady state. For the case of Cr, Ni, and Co coated STO, meaningful reconstructions of STO were still possible since the micro-fractures did not occur too often. Out of 18 datasets with micro-fractures, 16 have volumes of more than three million counts between micro-fracture events, which allowed for an accurate determination of the composition and analysis of trace elements.

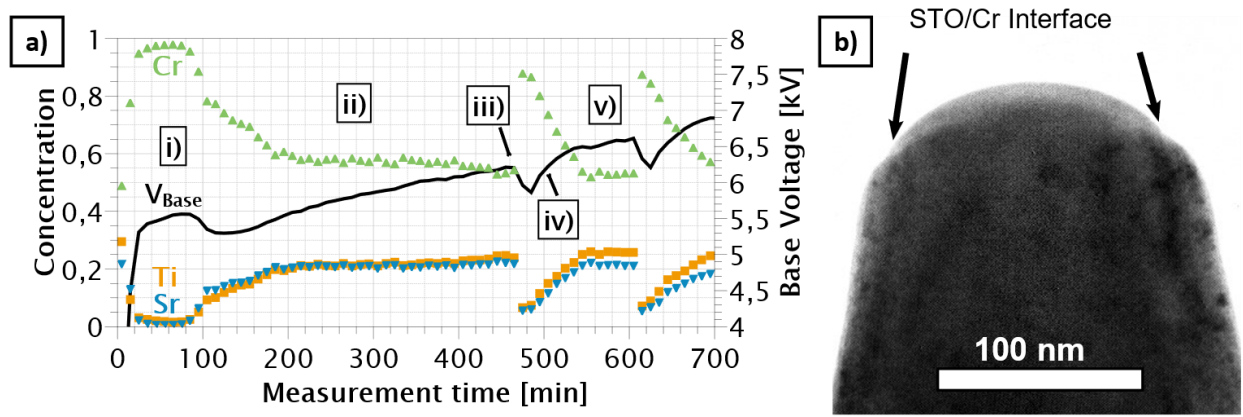

**Fig. 2.** *Micro-fractures during APT measurement of a Cr coated STO specimen.* **a)** *Voltage curve (black) and relative detection rates for Cr, Sr and Ti as a function of measurement time.* **b)** *TEM image of an APT specimen for which the measurement was interrupted during steady state field evaporation.*

**Mass Spectra, Reconstructions and Composition of Nb Doped, Cr Coated STO**

A representative mass spectrum from a Cr-coated 1 at% Nb-doped STO specimen is shown in Figure 3. Included is only the STO core region of the dataset and most of the Cr coating was removed by compositional filtering as detailed in supplementary note S2. Ti is present in various titanium oxide molecular ions with the main peak as $TiO^{2+}$, while Sr is only present as $Sr^{2+}$. Nb is present as $NbO_2^{1+}$ and $NbO_2^{2+}$. Measurements of Nb metal and Nb oxide specimens show that $Nb^{3+}$, $Nb^{2+}$ and $NbO^{2+}$ can also form sizable peaks at 31 Da, 46.5 Da and 70.5 Da (see Table S3) but are not visible in Figure 3. $Nb^{3+}$ may be present but has an overlap with $^{46}TiO^{2+}$ and would be hidden. We associate the peaks at 21 Da, 28 Da, and 29 Da with the residual gas in the Göttingen atom probe, since they did not appear in STO measurements performed on other systems. Except for $Sr^{2+}$, all peaks are sharp with slight thermal tailing and a mass resolution between $\frac{m}{\Delta m} = 300$ and $350$. The $Sr^{2+}$ peak has a lower mass resolution of $\frac{m}{\Delta m} \approx 40$ due to a stronger thermal tail and a secondary hump behind the main peak (see inset in Figure 3) that is mostly made up of single events (see Figure S4). The stronger thermal tail of $Sr^{2+}$ might be explained by a lower field evaporation threshold for Sr, as it could evaporate for a longer period of time after the laser pulse. The origins of the secondary hump are not clear. It is possible that the field evaporation of a different species leads to a local rearrangement of charges as was observed in FIM (Katnagallu et al., 2018). This would then leave Sr in a higher local field and lead to delayed field evaporation. Another possible explanation would be the dissociation of a larger molecular ion containing Sr shortly after field evaporation (Saxey, 2011).  However, unless the second species involved in both processes is neutral, the hump would show up in multi hit mass spectra and mass correlation plots of multi hit events, which it does not (Figure S4).

The main difference between the mass spectra of Nb doped STO and undoped STO is the presence of $NbO_2$ peaks (Figure S5 and Table S6). However, there are slight differences in the normalized peak heights and the doped mass spectrum has a higher background level and longer thermal tails. We attribute these to differences in the experimental conditions and specimen shape.

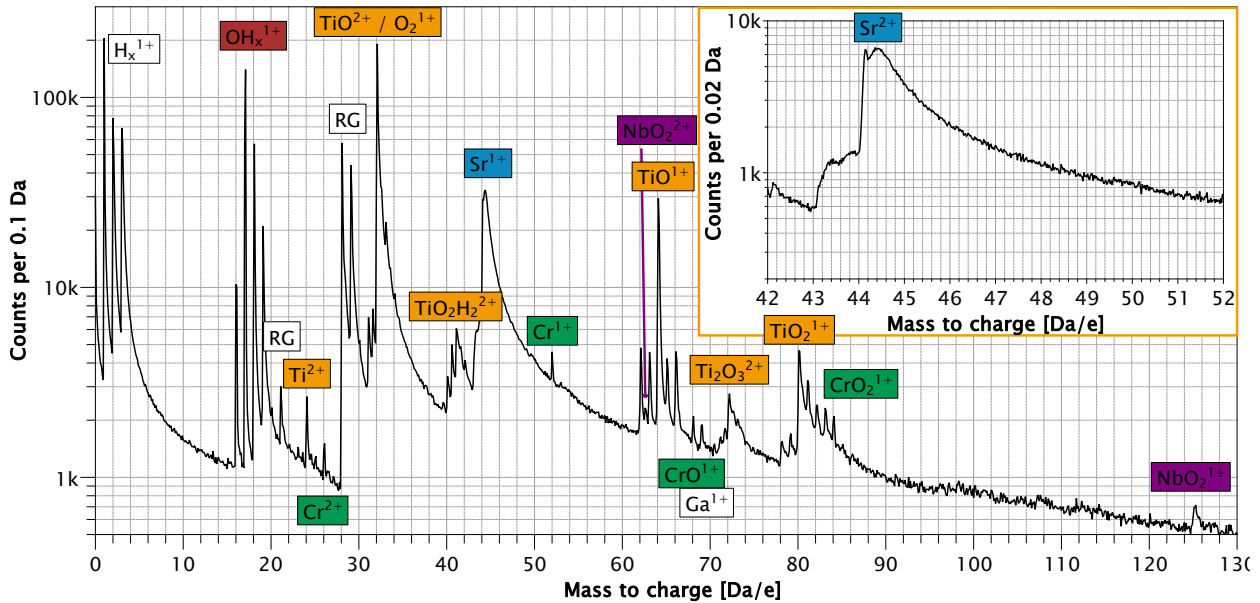

**Fig. 3.** *Representative mass spectrum from the core region of a Cr-coated 1 at% Nb-doped STO specimen. Counts from the coating have been removed by compositional filtering. A close up of the $Sr^{2+}$ peak shows the secondary hump.*

A full reconstruction of a Cr-coated, undoped STO specimen is shown in Figure 4a). The Cr coating (red) is clearly visible as an outer layer which surrounds the inner core of STO (blue). Concentration profiles over the Cr-STO interface (Figure 4b) show a thin layer with an increased proportion of $CrO^{1+/2+}$ ions followed by a layer with a higher Sr content compared to Ti and a higher total atomic density. It is unclear whether the Cr-O layer forms during specimen preparation or is a result of the formation of Cr-O ions during field evaporation. The layer of higher atomic density is also clearly visible in detector hit maps and most likely a result of the local magnification effect at interfaces between materials of different evaporation fields (Miller & Hetherington, 1991). This effect is present in measurements of specimen with Cr, Ni and Co coatings and is an indication that STO has a higher field evaporation threshold than the coating metals. The increase in Sr to Ti content at the interface might be explained as an artifact of the peak overlap between $^{48}$TiO$^{2+}$ and $^{16}$O$_2^+$ at 32 Da. Comparing Sr only with unaffected Ti peaks such as TiO$^{1+}$ shows no deviation of the Sr to Ti ratio at the Cr/STO interface. The $^{16}$O$_2^+$ contribution to the 32 Da peak as determined by comparison with $^{46}$TiO$^{2+}$ and $^{47}$TiO$^{2+}$ peaks reveal a higher $^{16}$O$_2^+$ contribution in the STO core (Figure S7).

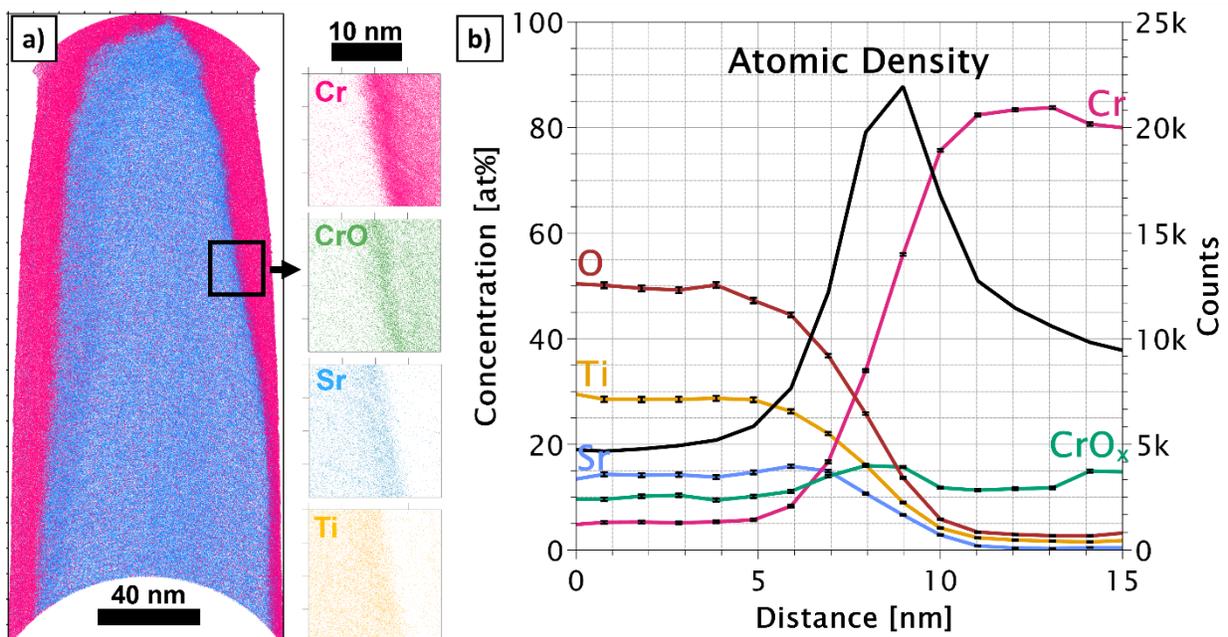

**Fig. 4.** *Representative reconstruction of a Cr-coated STO specimen* **a)** *A 20 nm thick slice through an APT reconstruction with a close-up view of the interface between the STO core and Cr-coating.* **b)** *Overall counts (black) as well as concentrations for Sr (blue), Ti (orange), Cr (pink) and CrO$_x$ (green) across the Cr/STO-interface. The Cr and CrO$_x$ left of the interface in the STO core are contributions from the background.*

The average composition of the STO core as determined from the mass spectrum in Figure 3 with conventional mass ranging is Sr: 17.90(2) at%, Ti: 26.90(2) at%, O: 55.20(2) at%. Compared to the nominal composition of Sr: 20 at%, Ti: 20 at%, O: 60 at%, this result is deficient in Sr and O. The oxygen content in laser-assisted APT measurements of insulating materials is known to vary with measurement conditions and is often underestimated due to the formation of neutral O species, that are not detected (Devaraj et al., 2013; Gault et al., 2016; Morris et al., 2024). The Sr deficit can be recovered, by extending the mass ranges to include the thermal tail and background behind the peak (Torkornoo et al., 2024). With the mass ranges for $TiO^{2+}$ and $Sr^{2+}$ extended (see Figure S8), the STO specimen from Figure 1 reaches a composition of Sr: 27.2 at%, Ti: 29.1 at% O: 43.7 at% with a Sr to Ti ratio of 0.93, close to the expected Sr to Ti ratio of 1. Composition profiles taken along the in-depth direction of a Nb-doped STO specimen reveal that the material is compositionally homogeneous (Figure 5); undoped specimens are similarly homogeneous. A FIB-induced Sr enrichment at the apex of the specimens as reported by Rodenbücher et al. (Rodenbücher et al., 2013) could not be observed.

The average Nb concentration is 0.600(3) at% compared to a nominal Nb concentration of 1 at%. This underestimation might be partially explained by an overlap of $Nb^{3+}$, a common Nb species, with $^{46}TiO^{2+}$. Any Nb arriving as $Nb^{3+}$ would be hidden by the overlap and not counted towards the Nb content. It is also possible, that some Nb is lost to DC field evaporation. Furthermore, the exact concentration of Nb is not provided by the manufacturer and might be lower than 1 at%.

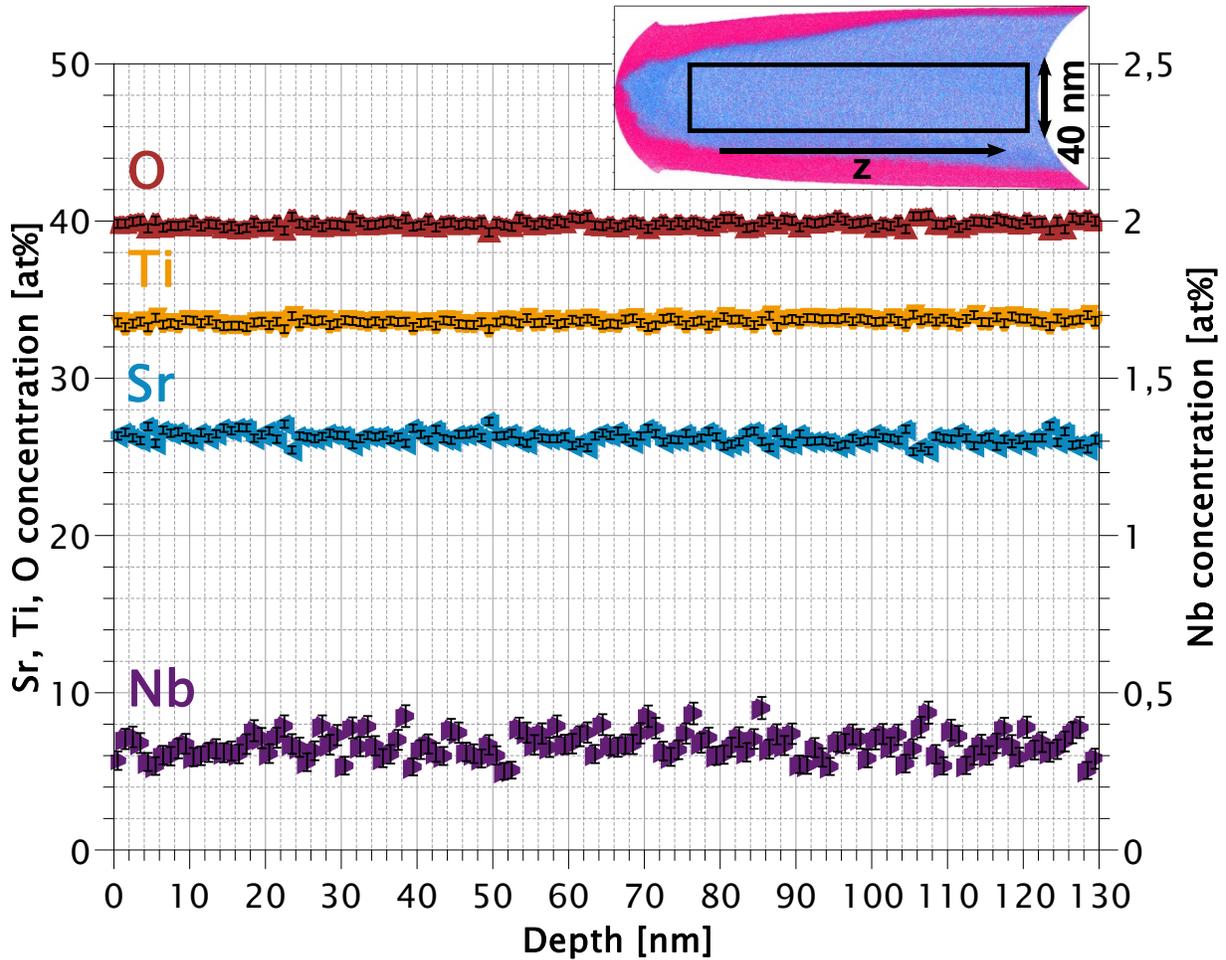

**Fig. 5.** *Composition profiles obtained from a Cr-coated 1 at% Nb-doped STO specimen. The data was extracted along the in-depth direction from the reconstruction shown in Figure 4a) using a 20 nm radius cylinder (see small inset) and a 0.1 nm bin size. Error bars were calculated according to (Danoix, et al., 2007). Sr, Ti, and O are homogeneously distributed within +/- 1 at% and Nb within +/- 10 at%. Note that the peak overlap of $^{48}TiO^{2+}$ and $^{16}O_2^{+}$ at 32 Da is not resolved here due to the inherent limitations of the profiling.*

## Discussion:

For specimens coated with Cr, Co or Ni, data acquisition at relatively high temperatures (> 115 K) and low detection rates (0.15 – 0.3 ions per 100 pulses) resulted in almost 100% successful measurements. Specimens measured under optimized conditions reveal uniform compositions, but with O and Nb deficiencies. Other than the presence of Nb, no significant differences between Nb doped and undoped STO was observed. Measurements of both materials were occasionally punctuated by micro-fractures, which introduced difficulties in building reliable reconstructions.

**Understanding Yield Improvements from Metal Coating and Measurement Conditions**

Brittle fracture of atom probe specimens is usually attributed to mechanical stresses induced by the electric field (Wilkes et al., 1972; Moy et al., 2011). The observed improvements in measurement success of STO from metal coating, high base temperatures and low evaporation rates can all be explained by a reduction in the electric field and field related mechanical stresses.

There are two straight-forward ways that the metal coating may protect against failure: by providing mechanical strengthening of the specimen and by shielding the STO material from the electric fields that induce mechanical stresses. STO exhibits brittle fracture in response to tensile stresses (Gumbsch et al., 2001). The metal coatings may be able to delay the fracture of the STO through their plasticity but will not significantly lower the stresses and strains in the STO by load sharing since they all have smaller moduli and strengths than the STO. Thus, mechanical contributions from the metal coatings seem unlikely to account for the improved performance of Cr, Co, and Ni coated specimens, nor for the poor performance of W coatings.

The extent that the metal coatings shield the STO from the applied electric fields depends strongly on how much of the STO is covered by the coating, as well as weakly on coating conductivity. The fields that reach the STO will produce mechanical stresses by both Coulomb forces between surface charges (Maxwell stresses) and by displacement of ions in the crystal lattice due to inverse piezoelectricity and electrostriction. The conductivities of the metal coating materials vary only slightly (with W being the highest), however the evaporation fields vary considerably, with that of W being twice as large as that of Cr (Cr: 27 V/nm, Ni: 35 V/nm, Co: 37V/nm, W: 52 V/nm) (Tsong, 1978), so that the metals will evaporate at different rates which will affect the evolution of the composite tip shape as will be discussed later. Metal coatings with evaporation fields much lower than of STO (we estimate the evaporation field of STO to be higher than Co based on the local magnification artifact), will be removed quickly, leaving exposed STO that may then fracture. On the other hand, coatings with larger evaporation fields will require larger applied electric fields to reach a given detection rate, possibly also leading to fracture of the STO. Presumably, the measurement conditions for the low evaporation fields of Cr, Co, and Ni relative to STO result in lower electric fields and relatively lower mechanical stresses in the STO during measurement.

The increased yield of coated STO specimen with higher base temperature and low field evaporation rates is likely due to the smaller electric field needed for evaporation at higher temperatures and low rates (Kelly & Larson, 2012). Maintaining a lower electric field reduces the mechanical stress on the specimen and lowers the risk of fracture. Furthermore, the temperature dependences of the dielectric constant and coefficients that couple strains to electric field (inverse piezoelectricity and electrostriction) may also play a role.

## Micro-fractures in Metal Coated STO Specimens

Micro-fractures are often observed for APT of brittle or heterogeneous materials, see (Yoon et al., 2008) for an example. In our case, we believe that small volumes of the STO material break off (Figure 2) as a result of a higher evaporation threshold for STO compared to the coating metals, leading to the scenario depicted in Figure 6. Initially, the entire tip is covered with metal and only ions from this coating are detected (Figure 6-i). When STO gets uncovered at the apex, the coating continues to evaporate preferentially until a steady state shape is reached, when the ratio between the evaporation rate of the coating and the STO core is constant (Figure 6-ii). A micro-fracture event then occurs when the mechanical stress in the exposed STO volume gets too large and the STO breaks off (Figure 6-iii). The fracture surface may be jagged, but any sharp protrusions quickly flatten out from field evaporation. The resulting specimen shape would then be flat in the STO region and highly curved at the edges, reducing the field and evaporation rate in the STO and increasing the field evaporation rate of the coating (Figure 6-iv). The metal continues to evaporate more quickly than the STO, until the specimen shape has reached the steady state shape again (Figure 6-v). We confirmed the proposed steady state shape via TEM imaging of a specimen where the APT measurement was interrupted during steady state evaporation (Figure 2b). A similar evolution with exposed core material and a coating was observed by Seol et al. (Seol et al., 2016) during field evaporation of Ag coated MgO.

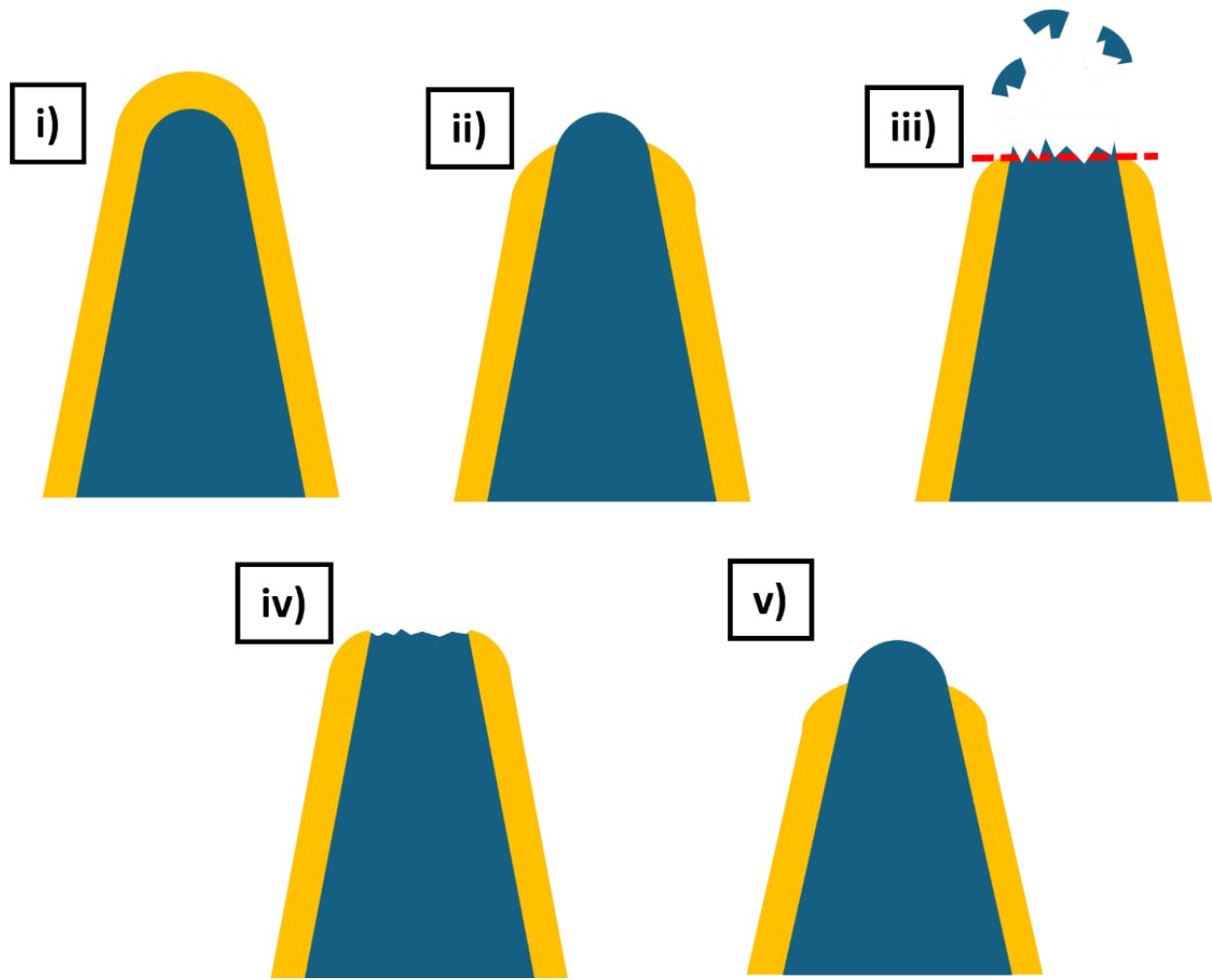

**Fig. 6.** *Sketch of the proposed evolution of the specimen shape: **i)** Initial state, **ii)** steady state shape, **iii)** micro-fracture, **iv)** recovery state, **v)** steady state.*

Similar evaporation behavior has been reported for W-coated BaTiO₃ (BTO) specimens (Jang et al., 2024). These specimens exhibited spikes in evaporation rate, comparable in magnitude to the spikes we observed during micro-fractures in STO. However, unlike the coated STO, where STO is detected at a reduced rate even directly after a micro-fracture, W coated BTO shows alternating field evaporation of W and BTO. As a result, the reconstructed datasets showed a layered specimen with alternating layers of W and BTO as opposed to the actual geometry of a continuous BTO core with an outer layer of W.

The micro-fractures have detrimental effects on the quality of the volume reconstructions. Since an unknown amount of material breaks off during each micro-fracture event, comparing with TEM images pre or post-measurement is no longer useful for calibrating depth or locating features of interest (see Fig. 3a in Arslan et al.) (Arslan et al., 2008). Furthermore, the change in electric field after a micro-fracture will produce deviations from the steady state composition, primarily through the effect of electric field on oxygen detection (Gault et al., 2016; Morris et al., 2019). Thus, if micro-fractures occur too frequently, it may be difficult to accurately determine the composition of the material. Moreover, depending on the microstructure and morphology of the specimen, features of interest like interfaces or small precipitates may be lost during a micro-fracture event. In the case of interfaces, this can be avoided by preparing specimens having the interface parallel to the specimen's main axis.

**Detectability of Trace Elements Exemplified by Nb in STO**

A criterion for the detectability of an APT peak above background level has been suggested by Pareige et al. (Pareige et al., 2016):

$$N_P - \alpha \cdot \sqrt{N_P \cdot (1-Q)} > N_B + \alpha \cdot \sqrt{N_B \cdot (1-Q)} \qquad 1$$

where $N_P$ is the peak height in counts, $N_B$ the local background in counts, $Q$ detection efficiency of the used atom probe and $\alpha$ the significance level. This criterion is equivalent to requiring that the peak height be larger than the background plus $\alpha$ times the standard deviation, which is estimated by the square root term. Pareige et al. suggest using a value $\alpha = 1$; after examining our mass spectra (see Figure S9), we chose a slightly more rigorous criterion of $\alpha = 2$ for detecting the peak of interest (NbO$_2^{1+}$).

For assessing detectability limits, we consider a homogeneous volume of material $V_o$, containing a given total number of atoms $N_o$, and a certain number of Nb atoms $x \cdot N_o$, where $x$ is the atomic fraction of Nb atoms in the volume. We then expect $Q \cdot x \cdot N_o$ impurity atoms to arrive at the detector in various molecular and ionization states. The different states contribute to different peaks in the spectrum. Assuming a single peak is used to assess detectability, the fraction of impurities arriving at the detector that contribute to the peak of interest is defined as $f$. Thus, the number of impurity atoms in the peak of interest is $f \cdot Q \cdot x \cdot N_o$. We further make the simplifying assumption that the peak height in counts $N_P$ is simply proportional to the number of ions detected in the given peak via a proportionality constant $c_P$ with units of counts/atom, so that

$$N_P = c_P \cdot f \cdot Q \cdot x \cdot N_o \qquad\qquad 2$$

Analogously, we assume that the local background height in counts can be estimated as

$$N_B = c_B \cdot Q \cdot N_o \qquad\qquad 3$$

thereby assuming that the background at the peak of interest is simply proportional to the total number of ions that arrive at the detector.

We apply this criterion to assess the detectability of Nb in Nb doped STO. The composition profile of Nb shown in Figure 5 and investigations of the Nb distribution in Verneuil-grown STO (Rodenbücher et al., 2016) suggest, that Nb is distributed homogeneously in 1 at% Nb doped STO crystals. The values for $c_P$ and $c_B$ can be determined using Equations (2) and (3) and the spectrum for 1 at% Nb in STO in Figure 1. We further use a value of $Q = 0.4$ (Göttingen atom probe) and estimate $f = 0.3$ for the NbO$_2^{1+}$ peak, based on the ratio of NbO$_2^{1+}$ to all visible Nb peaks in Figure 1. We then use Equation (1) to calculate the detection limit of Nb in STO as a function of the analyzed volume using the NbO$_2^{1+}$ peak (Figure 7). In order to detect 1 at% Nb, an STO volume with a radius of 16 nm (19200 nm$^3$) containing $1.61 \cdot 10^6$ atoms, must be evaporated. For a larger analyzed volume, with a radius of around 50 nm (containing $\sim 4.7 \cdot 10^7$ atoms), the minimum detectable concentration of Nb in STO is 0.75 at%, while in an analyzed volume with a radius of 3 nm (containing $\sim 8000$ atoms), the minimum detectable concentration is 10 at%. We note that because the background is assumed to be independent of the material composition in the simple approximation presented here, this approach is best used for dilute impurity concentrations that do not deviate much from the reference measurement.

One of the main applications of APT is to detect nanoscale variations in composition. The detection limit presented here, which gives the minimum impurity concentration that can be detected when analyzing a given homogeneous volume of material, can also be used to estimate the spatial and concentration detection limits for heterogeneous materials. To first order, the data shown in Figure 7 can therefore be used to estimate the minimum impurity composition that can be detected in a cluster of a given size. So, for example, if Nb were distributed inhomogeneously in a STO matrix, our measurement method would be able to detect 2 at% Nb clusters with radii as small as 6 nm. The smaller the cluster size, the larger the impurity concentration must be to be detectable.

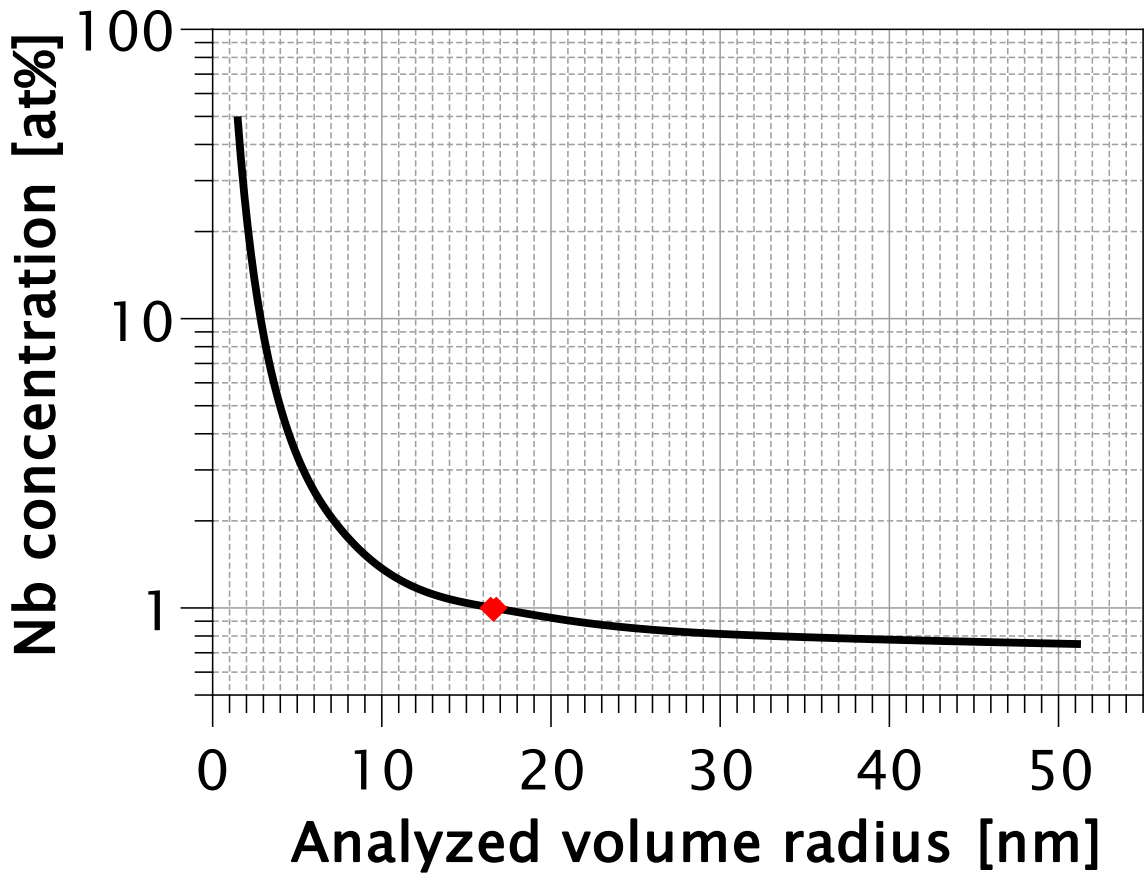

**Fig. 7.** *Detection limit of Nb in STO using the NbO$_2^{1+}$ peak, as a function of the cluster or analyzed volume radius. STO volumes with 1 at% Nb (red point) can be detected if the volume has a radius larger than 16 nm. Larger volumes are needed to detect smaller concentrations.*

The detection limits calculated here for Nb are less sensitive than some of the limits that have been discussed in the literature (e.g. 0.02 at% for Mn in BTO (Jang et al., 2024)). Detection limits depend sensitively on the material system and impurity of interest, as well as on the measurement system. In particular, impurities that only form one peak in the mass spectrum are expected to have a more sensitive detection limit. Further, the local background level has a strong influence on the detectability of a given peak. Our measurements of STO show a high background compared to measurements of Al, Fe, and W in the Göttingen atom probe, likely due to the high base temperature and low evaporation rate required for a good measurement yield. Additional system parameters that influence the detectability of small concentrations are background contributions from electronic noise and residual gas in the system as well as the mass resolution. Atom probes with a higher mass resolution will produce sharper peaks in the mass spectrum which are easier to detect over noise in the background.

## Conclusion:

This study establishes a robust APT methodology for nanoscale chemical analysis of strontium titanate. By applying thin metal coatings to STO specimens, high-yield APT measurements become feasible, since the coatings shield the material from strong electric fields and reduce field-induced mechanical stress and fracture. Importantly, no difference in APT measurement success was observed between doped and undoped STO samples, indicating that the intrinsic conductivity of the specimens becomes negligible once coated. The detection limit for Nb as determined using the $NbO_2^{1+}$ peak, is approximately 0.7 at% for a homogenous solution. However, identifying Nb within clusters requires higher local concentrations. The detection limits for other impurities in STO, particularly those that produce single peaks in the mass spectrum, are expected to be lower.

**Acknowledgements:**

This research is funded by the German Science Foundation (DFG) through the research unit FOR 5065 "Energy Landscapes and Structure in Ion Conducting Solids" (ELSICS), project number 428906592. Access to equipment of the "Collaborative Laboratory and User Facility for Electron Microscopy" (CLUE) at University of Göttingen and to the atom probe facilities at the Institute for Materials Science, Stuttgart under Prof. Dr. Dr. h.c. Guido Schmitz are gratefully acknowledged. We would like to thank Tim M. Schwarz from Max-Planck-Institute for Sustainable Materials, Düsseldorf, for preliminary measurements of uncoated STO.

**Supplementary information**

| Specimen: | Measured at: | Coating: | Approximate Counts before termination: | Reson for termination: | Detection rate target [ions per 100 pulses]: | T$_{base}$ [K] | LPE [µJ] ** |
|---|---|---|---|---|---|---|---|
| STNO 1 | IMW Stuttgart | - | 6000 | fracture | 0.2 - 0.3 | 70 | 0.06 |
| STNO 2 | IMW Stuttgart | - | 650000 | fracture | 0.1 - 0.15 | 70 | 0.05 |
| STNO 3 | IMW Stuttgart | - | 850000 | fracture | 0.3 - 0.6 | 70 | 0.08 |
| STNO 4 | IMW Stuttgart | - | 370000 | fracture | 0.2 - 0.3 | 70 | 0.06 |
| STNO 5 | IMW Stuttgart | - | 100000 | fracture | 0.2 - 0.3 | 210 | 0.06 |
| STNO 6 | IMW Stuttgart | - | 15000 | fracture | 0.2 - 0.3 | 210 | 0.06 |
| STNO 7 | IMW Stuttgart | - | 100000 | fracture | 0.2 - 0.3 | 210 | 0.06 |
| STNO 8 | IMW Stuttgart | - | 180000 | fracture | 0.2 - 0.3 | 210 | 0.06 |
| STNO 9 | IMW Stuttgart | - | 160000 | fracture | 0.2 - 0.3 | 210 | 0.06 |
| STNO 10 | IMW Stuttgart | - | 580000 | fracture | 0.2 - 0.3 | 210 | 0.06 |
| STNO 11 | IMW Stuttgart | - | 90000 | fracture | 0.2 - 0.3 | 210 | 0.06 |
| STNO 12 | IMW Stuttgart | - | 750000 | fracture | 0.2 - 0.3 | 210 | 0.06 |
| STNO 13 | IMW Stuttgart | - | 550000 | fracture | 0.3 - 0.6 | 210 | 0.06 |
| STNO 14 | IMW Stuttgart | - | 160000 | fracture | 0.3 - 0.6 | 210 | 0.06 |
| STNO 15 | IMW Stuttgart | - | 300000 | fracture | 0.3 - 0.6 | 210 | 0.074 |
| STNO 16 | IMW Stuttgart | - | 100000 | fracture | 0.3 - 0.6 | 210 | 0.075 |
| STNO 17 | IMP Göttingen | - | 340000 | fracture | 0.15 - 0.3 | 120 | 0.07 |
| STNO 18 | IMP Göttingen | - | 560000 | fracture | 0.15 - 0.3 | 120 | 0.075 |
| STNO 19 | IMP Göttingen | Cr | 2000000 | fracture | 0.15 - 0.3 | 70 | 0.092 |
| STNO 20 | IMP Göttingen | Cr | 3300000 | fracture | 0.15 - 0.3 | 60 | 0.06 |
| STNO 21 | IMP Göttingen | Cr | 3100000 | fracture | 0.2 - 0.4 | 70 | 0.1 |
| STNO 22 | IMP Göttingen | Cr | 11500000 | fracture | 0.15 - 0.3 | 115 | 0.074 |
| STNO 23 | IMP Göttingen | Cr | 8400000 | fracture | 0.15 - 0.3 | 120 | 0.113 |
| STNO 24 | IMP Göttingen | Cr | 53000000 | other* | 0.15 - 0.3 | 120 | 0.12 |
| STO 1 | IMP Göttingen | Cr | 16000000 | max. voltage | 0.15 - 0.6 | 115 | 0.071 |
| STO 2 | IMP Göttingen | Cr | 13000000 | max. voltage | 0.15 - 0.3 | 120 | 0.073 |
| STO 3 | IMP Göttingen | Cr | 42000000 | max. voltage | 0.15 - 0.3 | 120 | 0.118 |
| STO 4 | IMW Stuttgart | Cr | 15000000 | fracture | 0.15 - 0.3 | 100 | 0.125 |
| STO 5 | IMP Göttingen | Cr | 6000000 | max. voltage | 0.15 - 0.3 | 120 | 0.0725 |
| STO 6 | IMP Göttingen | Cr | 16000000 | other* | 0.15 - 0.3 | 120 | 0.095 |

| STO 7 | IMP Göttingen | Cr | 14800000 | other* | 0.15 - 0.3 | 120 | 0.09 |
| STO 8 | IMP Göttingen | Cr | 60000000 | other* | 0.15 - 0.3 | 120 | 0.113 |
| STO 9 | IMP Göttingen | Cr | | fracture | 0.2 - 0.4 | 120 | 0.108 |
| STO 10 | IMP Göttingen | Ni | 17000000 | fracture | 0.15 - 0.3 | 120 | 0.108 |
| STO 11 | IMP Göttingen | Co | 6900000 | fracture | 0.15 - 0.3 | 120 | 0.115 |
| STO 12 | IMP Göttingen | Co | 31500000 | fracture | 0.15 - 0.3 | 120 | 0.11 |
| STO 13 | IMP Göttingen | Cr | 60000000 | other* | 0.15 - 0.3 | 120 | 0.11 |
| STO 14 | IMP Göttingen | Cr | 11000000 | fracture | 0.15 - 0.3 | 120 | 0.105 |
| STO 15 | IMP Göttingen | Cr | 11000000 | fracture | 0.15 - 0.3 | 120 | 0.105 |
| STO 16 | IMP Göttingen | Cr | 30000000 | fracture | 0.15 - 0.3 | 90 | 0.106 |
| STO 17 | IMP Göttingen | Cr | 87000000 | other* | 0.15 - 0.3 | 120 | 0.112 |

**Supplementary Table S1:** *List of performed measurements. Measurements of STNO-1 to STNO-16 at the Institute for Materials Science (IMW) at the University of Stuttgart were performed by T. M. Schwarz.*

*\*other reasons for termination include interruption for TEM characterization and errors with the instrument.*

*\*\*due to differences in the laser setup, laser pulse energies (LPE) of measurements performed at IMW Stuttgart and IMP Göttingen are not comparable.*

**Supplementary Note S2:** *For the mass spectrum analysis, counts from the metal coatings were removed by compositional filtering with a custom python script. The dataset was subdivided into 1 nm³ sized cube voxels and the fraction of all coating related ions (e.g. for Cr coating: Cr, $CrO^{1+/2+}$, $CrO_2^{1+}$, $Cr_2O^{2+}$) was calculated. Voxels with more than 30% coating related counts and voxels with less than 10 overall counts at the edge of the dataset were* *then dropped from further analysis. Other values for the voxel size and filtering threshold were tested and 1 nm³ voxels with a 30% threshold were found to work best.*

| Ion: | Mass to charge [Da/e]: | Nb oxide rel. peak heights: | Nb metal rel. peak heights: | Nb in STO rel. peak heights |
|---|---|---|---|---|
| Nb$^{3+}$ | 31 | -/- | 1 | -/- |
| NbO$^{3+}$ | 36.33 | 0.019 | 0.013 | -/- |
| Nb$^{2+}$ | 46.5 | 0.009 | 0.75 | -/- |
| NbO$^{2+}$ | 54.5 | 0.5 | 0.1 | -/- |
| NbO$_2^{2+}$ | 62.5 | 1 | 0.004 | 1 |
| NbO$_3^{2+}$ | 70.5 | 0.025 | -/- | -/- |
| Nb$^{1+}$ | 93 | -/- | -/- | -/- |
| NbO$^{1+}$ | 109 | 0.002 | -/- | -/- |
| NbO$_2^{1+}$ | 125 | 0.14 | 0.004 | 0.43 |
| NbO$_3^{1+}$ | 141 | 0.004 | -/- | -/- |
| NbO$_4^{1+}$ | 157 | 0.002 | -/- | -/- |

**Supplementary Table S3:** *List of mass to charge ratios and relative peak heights for Nb containing peaks obtained from APT measurements of Nb metal, Nb oxide and Nb in STO.*

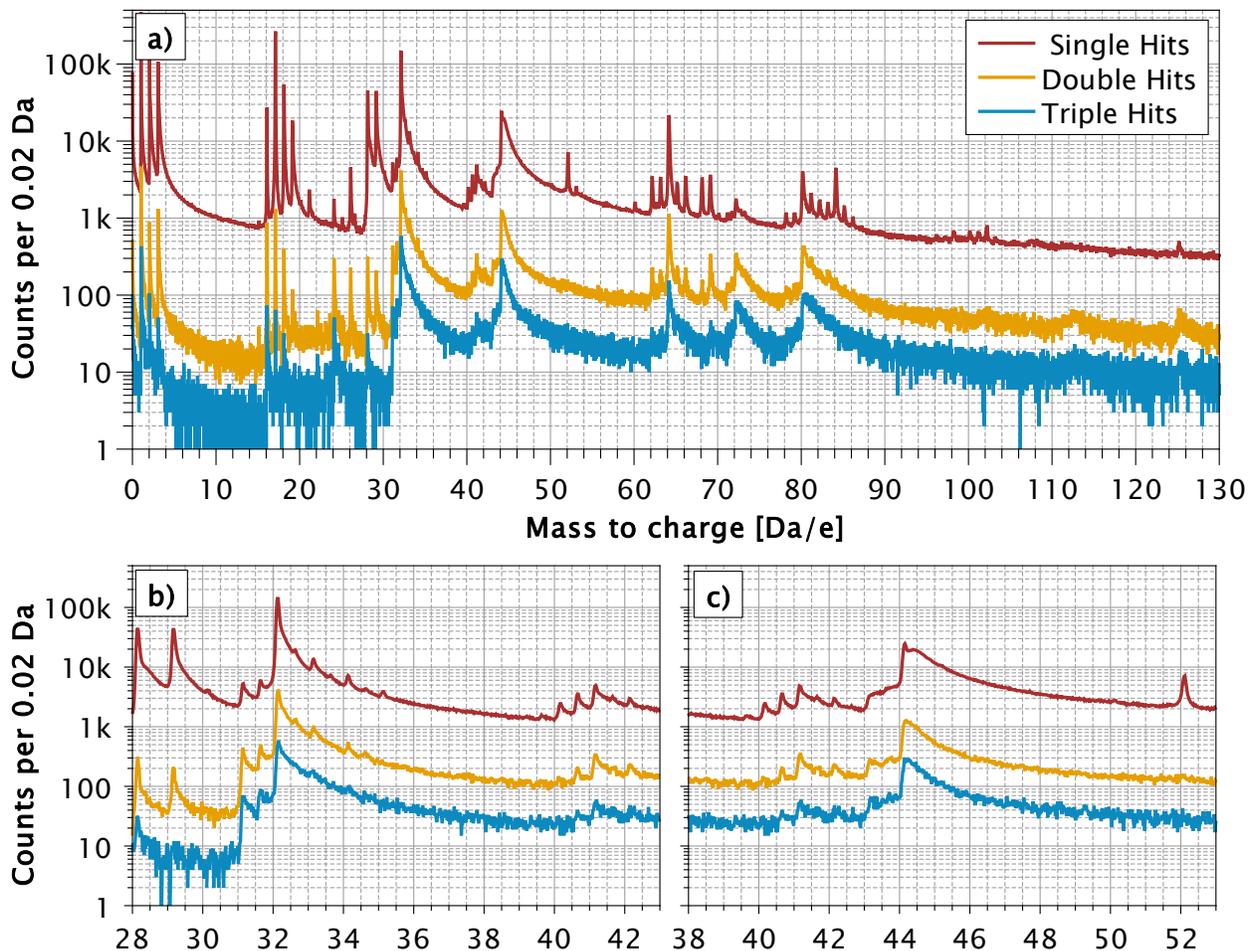

**Supplementary Figure S4:** *Mass spectra of single-, double- and triple hits of the specimen shown in Figure 1 **a)** Overall mass spectrum showing that residual gas peaks are less prominent in the double and triple hit spectra. **b)** Close up of the TiO²⁺ peak. The main peak at 32 Da is less prominent in the double and triple hit spectra compared to the isotope peaks at 31 and 31.5 Da and the thermal tail behind 32 Da.  **c)** Close up of the Sr²⁺ peak. The secondary hump originates mostly from single hits.*

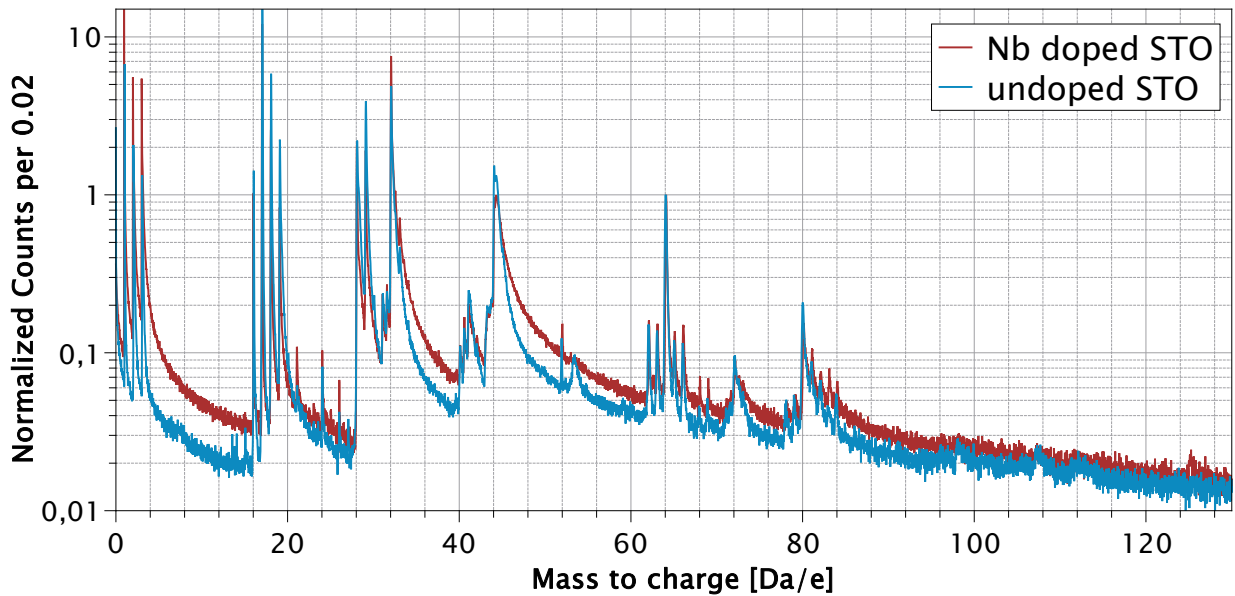

**Supplementary Figure S5:** *Mass spectra comparison between undoped STO (blue) and 1at% Nb doped STO (red).*

| Ion: | Mass to charge: | Rel. peak height undoped STO: | Rel. peak height Nb doped STO |
|------|-----------------|-------------------------------|-------------------------------|
| H$^{1+}$ | *1* | *1.7* | *1* |
| O$^{1+}$ | *16* | *0.8* | *0.35* |
| $^{46}$TiO$^{2+}$ | *31* | *0.23* | *0.23* |
| $^{48}$TiO$^{2+}$/O$_2^+$ | *32* | *3.7* | *6.5* |
| TiO$_2$H$_2^{2+}$ | *42* | *0.11* | *0.14* |
| Sr$^{2+}$ | *44* | *1.35* | *1* |
| Sr$^{2+}$ hump | *44.3* | *1.36* | *1.1* |
| TiO$^{1+}$ | *64* | *1* | *1* |
| Ti$_2$O$_3^{2+}$ | *72* | *0.09* | *0.09* |
| TiO$_2^{1+}$ | *80* | *0.21* | *0.16* |

**Supplementary Table S6:** *Heights of selected peaks normalized to TiO$^{1+}$ for the mass spectra of undoped and 1 at% Nb doped STO shown in Figure S4.*

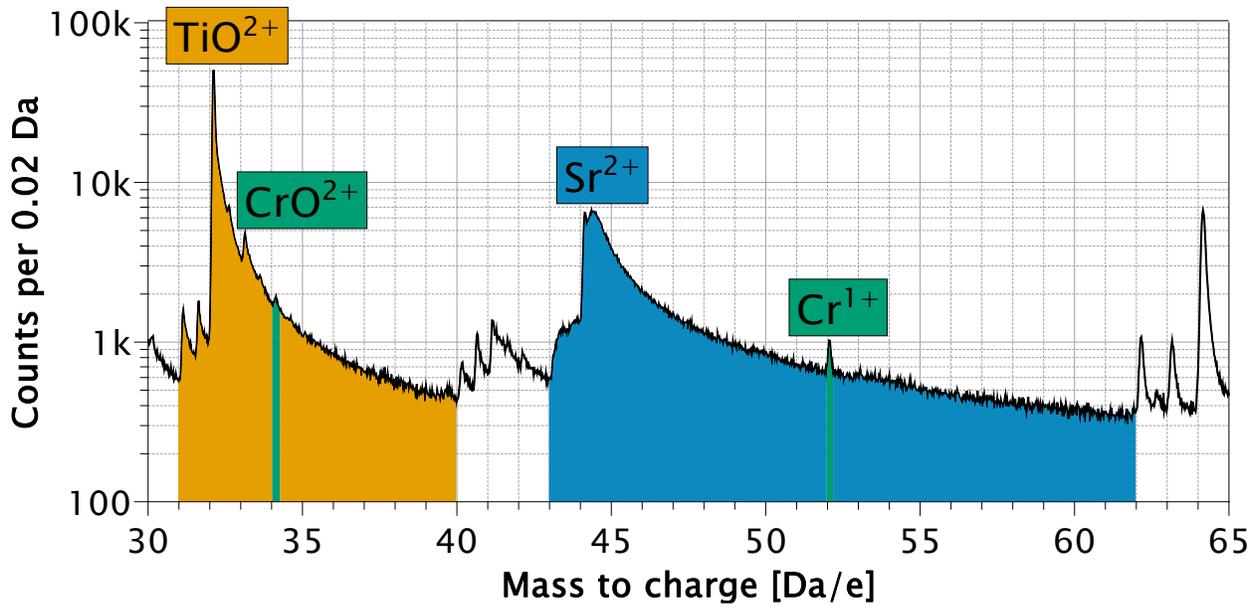

**Supplementary Figure S7:** *Modified mass ranges for TiO²⁺ and Sr²⁺.*

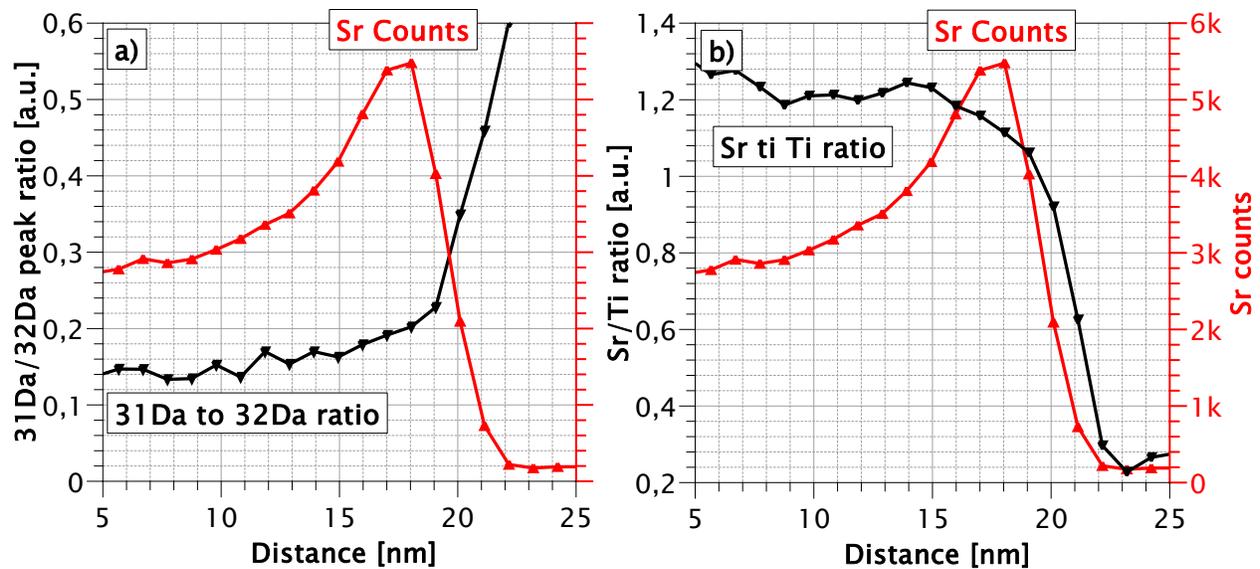

**Supplementary Figure S8:** *Interface between the STO core and Cr coating* **a)** *Ratios of ⁴⁶TiO²⁺ and ⁴⁷TiO²⁺ to ⁴⁸TiO²⁺/O₂¹⁺* **b)** *Sr/Ti ratio without TiO²⁺. Displayed in red are the absolute Sr counts to show the position of the interface. The profiles have a bin width of 1 nm.*

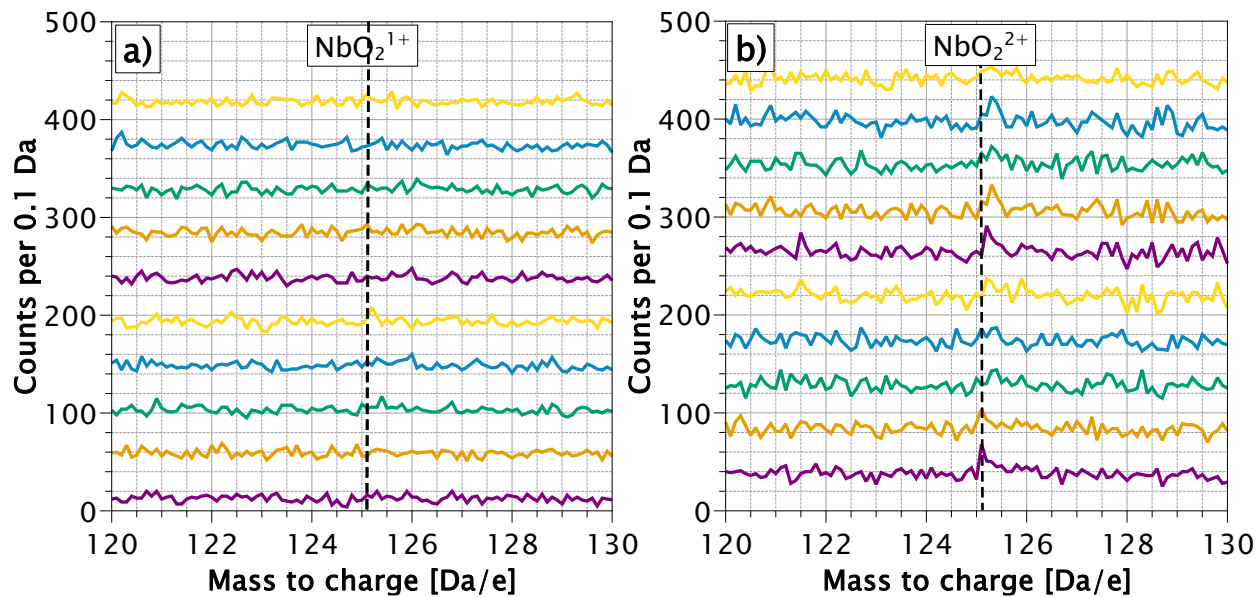

**Supplementary Figure S9:** *Stacked mass spectra from spherical sub volumes with **a)** a radius of 12 nm, satisfying Equation 1 with a significance of $\alpha = 1$ and **b)** a radius of 17 nm, satisfying Equation 1 with a significance of $\alpha = 2$.*